\documentclass[letter,twocolumn]{jpsj3}

\usepackage{ulem}
\usepackage{mathrsfs}
\usepackage{amsmath}
\usepackage{amsfonts}
\usepackage{amssymb}
\usepackage{txfonts}
\usepackage{bm}
\usepackage{tabularx}
\usepackage{graphicx,color}
\usepackage{cite}

\def\Vec#1{\bm{#1}}
\def\Hc2{H_\mathrm{c2}}

\author{
	Shunichiro Kittaka$^1$\thanks{kittaka@issp.u-tokyo.ac.jp}, 
	Takanori Taniguchi$^1$\thanks{Present address: Institute for Materials Research, Tohoku University, Sendai 980-8577, Japan}, 
	Kazumasa Hattori$^2$,
	Shota Nakamura$^{1}$\thanks{Present address: Department of Physical Science and Engineering, Nagoya Institute of Technology, Nagoya 466-8555, Japan},\\
	Toshiro Sakakibara$^{1}$,
	Masashi Takigawa$^{1}$,
	Masaki Tsujimoto$^{1\thanks{Present address: Asahi Kasei Microdevices Corporation}}$,
	Akito Sakai$^{1}$,\\
	Yosuke Matsumoto$^{1}$\thanks{Present address: Max Planck Institute for Solid State Research, Heisenbergstrasse 1, 70569 Stuttgart, Germany}, and
	Satoru Nakatsuji$^{1,3}$
}

\inst{$^{1}$Institute for Solid State Physics, University of Tokyo, Kashiwa, Chiba 277-8581, Japan\\
      $^{2}$Department of Physics, Tokyo Metropolitan University, Hachioji, Tokyo, 192-0937, Japan\\
      $^{3}$Department of Physics, University of Tokyo, Hongo, Bunkyo-ku, Tokyo 113-0033, Japan\\
}

\begin{document}

\title{Field-Orientation Effect on Ferro-Quadrupole Order in PrTi$_2$Al$_{20}$}

\date{\today}

\abst{
Ferro-quadrupole (FQ) order in the non-Kramers $\Gamma_3$ doublet system PrTi$_2$Al$_{20}$ 
has been investigated via angle-resolved measurements of the specific heat, rotational magnetocaloric effect, and entropy,
under a rotating magnetic field within the $(1\bar{1}0)$ plane.
The FQ transition occurring at 2~K is robust when the magnetic field $B$ is applied precisely along the $[111]$ direction or in low fields below 1~T.
By contrast, the magnetic field of larger than 1~T tilted away from the $[111]$ direction sensitively changes the FQ transition to a crossover.
The energy gap between the ground and first-excited states in the FQ order increases remarkably 
with the magnetic field in $B \parallel [001]$,
but hardly depends on the magnetic-field strength,
at least up to 5~T, in the field orientation between the $[111]$ and $[110]$ axes.
These features can be reproduced by using a phenomenological model for FQ order assuming an anisotropic field-dependent interaction between quadrupoles, 
which has been recently proposed to explain the field-induced first-order phase transition in PrTi$_2$Al$_{20}$.
The present study demonstrates the great potential of the field-angle-resolved measurements 
for evaluating possible scenarios for multipole orders.
}

\maketitle
Multipole degrees of freedom often play an essential role in causing a variety of ordered phases in spin-orbit coupled systems.
Recently, the Pr$T_2X_{20}$ system ($T$=transition elements, $X$=Al or Zn) with the cubic CeCr$_2$Al$_{20}$-type structure has attracted much attention 
because it has a non-Kramers $\Gamma_3$ doublet ground state under the crystalline electric field (CEF).\cite{Onimaru2016JPSJ} 
These $\Gamma_3$-doublet materials provide a significant chance to study novel phenomena 
caused by electric quadrupoles ($O_{20}$ and $O_{22}$) and a magnetic octupole ($T_{xyz}$) in the absence of active magnetic dipoles.
Furthermore, it provides a unique opportunity to investigate the relation between exotic superconductivity and multipole fluctuations because 
superconductivity likely driven by multipole fluctuations has been found in PrTi$_2$Al$_{20}$,\cite{Sakai2012JPSJ,Matsubayashi2012PRL} PrV$_2$Al$_{20}$,\cite{Tsujimoto2014PRL} PrIr$_2$Zn$_{20}$,\cite{Onimaru2010JPSJ} and PrRh$_2$Zn$_{20}$.\cite{Onimaru2012PRB}

In this letter, we focus on multipole order in PrTi$_2$Al$_{20}$ occurring at $T_{\rm FQ} \sim 2$~K.\cite{Sakai2011JPSJ,Ito2011JPSJ}
The ground state of the non-Kramers doublet $\Gamma_3$ and the first-excited state of the triplet $\Gamma_4$ are well separated by a substantial energy gap of 54~K.\cite{Sato2012PRB} 
Therefore, it is expected that 
the low-temperature physics in this material is dominated by degrees of freedom of the non-magnetic $\Gamma_3$ doublet. 
On the basis of broadening of the transition under a magnetic field $B$ parallel to the $[100]$ axis \cite{Sakai2011JPSJ} 
together with the absence of a magnetic superlattice reflection in the neutron scattering experiment,\cite{Sato2012PRB} 
the multiple order can be attributed to ferroquadrupole (FQ) one. 
Moreover, evidence for ferro order of $O_{20}$ quadrupole has been provided from NMR measurements.\cite{Taniguchi2016JPSJ}

Recently, $^{27}$Al-NMR and magnetization measurements have revealed highly-anisotropic phase diagram of the FQ order in PrTi$_2$Al$_{20}$.\cite{Taniguchi2019JPSJ}
A field-induced first-order phase transition was found in the FQ ordered state near 2~T for $B \parallel [100]$ and $[110]$,
whereas it was absent in $B \parallel [111]$.
Such a highly-anisotropic phase diagram is unexpected for the $\Gamma_3$-doublet system because the magnetic dipole is not active.\cite{Hattori2014JPSJ,Hattori2016JPSJ}
As a possible origin, competition between the Zeeman effect and multipole interactions was proposed in Ref.~\ref{Taniguchi2019S},
resulting in switching of the FQ order parameters under $B$.
Indeed, by assuming an anisotropic \textit{field-dependent} quadrupole-quadrupole interaction, which is overcome by the Zeeman effect in high fields, 
the phase diagrams for three field orientations, $B \parallel [100]$, $[111]$, and $[110]$, have been successfully reproduced
by a Landau theory based on the mean-field approximation.\cite{Taniguchi2019JPSJ} 
However, this scenario is based on a phenomenological assumption whose validity has not yet been established, e.g., from the microscopic theory. 
Further investigations are necessary to elucidate the FQ order parameters of PrTi$_2$Al$_{20}$, both experimentally and theoretically.

In this study, the field-orientation effect on this highly-anisotropic phase diagram of PrTi$_2$Al$_{20}$ has been investigated precisely from thermal experiments. 
The specific heat is particularly useful to distinguish between the low-field and field-induced FQ phases;
a sharp (broad) peak in the specific heat corresponds to an occurrence of the low-field (field-induced) FQ phase
because the FQ transition at $T_{\rm FQ}$ changes into a crossover in the field-induced phase due to the predominant Zeeman effect.
Our experimental results demonstrate that the broadening of the FQ transition is sensitively induced by a magnetic field of larger than 1~T applied away from the $[111]$ axis.
Furthermore, it has been clarified that, when the field angle is in the range between the $[111]$ and $[110]$ directions,
the energy gap between the ground and first-excited states in the FQ ordered state is nearly invariant with $B$, at least up to 5~T.
These experimental results can be explained by a phenomenological model including the field-dependent anisotropic quadrupole interaction, 
which was introduced in the previous study~\cite{Taniguchi2019JPSJ}.

Single crystals of PrTi$_2$Al$_{20}$ were grown by an Al self-flux method~\cite{Sakai2011JPSJ}. 
A single crystal with its mass of 1.258~mg was used in this study, having a small, triangular facet which corresponds to the $(111)$ plane.
The residual resistivity ratio of a crystal cut from the same batch is as large as 170, which ensures high quality of the present sample.
This $(111)$ sample face was attached on a sample stage of our home-made calorimeter~\cite{Kittaka2018JPSJ}, 
and it was placed in a dilution refrigerator so that the $[1\bar{1}0]$ axis of the sample is roughly parallel to the vertical $z$ direction.
A magnetic field was generated within the $xz$ plane by using a vector magnet, and 
the refrigerator was rotated around the $z$ axis by using a stepper motor which was mounted at the top of the dewar.
By using this system, the orientation of the magnetic field can be controlled three-dimensionally. 
In this study, the magnetic field was aligned precisely parallel to the $(1\bar{1}0)$ plane with an accuracy of better than $0.1^\circ$.
The field angle $\phi_B$ is defined as the angle between the $[001]$ axis and the magnetic field.

\begin{figure}
\begin{center}
\includegraphics[width=2.75in]{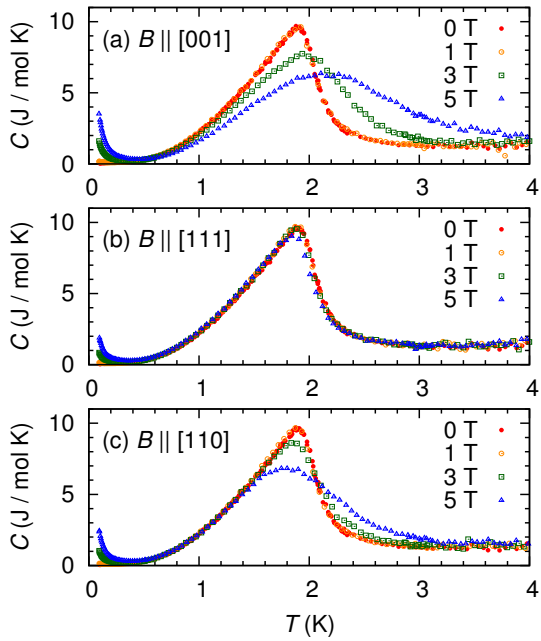}
\caption{
(Color online) Temperature dependence of the specific heat under a magnetic field parallel to the (a) $[001]$, (b) $[111]$, and (c) $[110]$ axes.
}
\label{CT}
\end{center}
\end{figure}

The specific heat $C$ and the rotational magnetocaloric effect $(\partial T/\partial\phi_B)_{S,B}$ were investigated by the standard semi-adiabatic method~\cite{Kittaka2018JPSJ}
in a rotating magnetic field within the $(1\bar{1}0)$ plane.
From the field-angle anisotropy in $C$,
it was found that the $[1\bar{1}0]$ axis was tilted away from the rotational axis ($\parallel z$) by approximately $10.6^\circ$.
Therefore, in all the field-angle-resolved measurements shown here, 
we took two processes to rotate the magnetic field within the $(1\bar{1}0)$ plane; 
first, the refrigerator was rotated to a target angle, and then, horizontal and vertical magnetic fields were fine-tuned 
so that the resulting field orientation is parallel to the $(1\bar{1}0)$ plane.
During the first process, we precisely measured the relative change in the sample temperature, $\delta T(\phi_B)=T(\phi_B)-T_0$, 
upon a semi-adiabatic small-angle rotation of the magnetic field (typically $1^\circ$).
Here, $T_0$ is the base temperature of the sample. 
The rotational magnetocaloric effect was evaluated from the initial slope of $\delta T(\phi_B)$ just after starting the field rotation. 
Then, the field-angle dependence of the entropy $S(\phi_B)$ can be obtained from the relation 
\begin{equation}
\biggl(\frac{\partial S}{\partial \phi_B}\biggl)_{T,B}=-\frac{C}{T}\biggl(\frac{\partial T}{\partial \phi_B}\biggl)_{S,B}.
\end{equation}

\begin{figure}
\begin{center}
\includegraphics[width=2.75in]{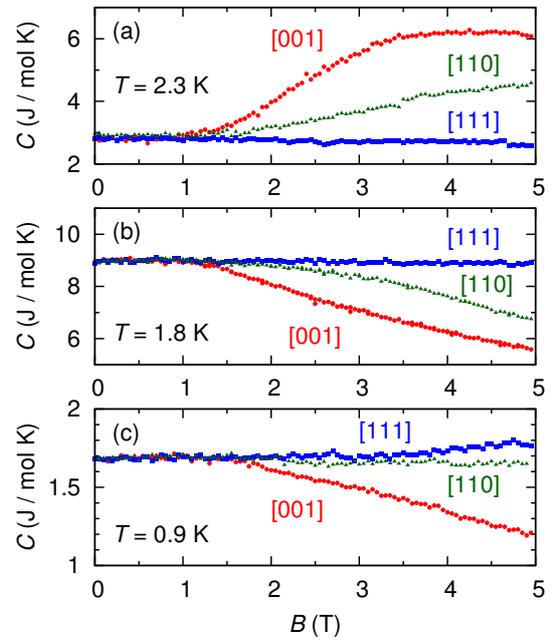}
\caption{
(Color online) Field variation of the specific heat at (a) 2.3, (b) 1.8, and (c) 0.9 K. Labels represent the field orientation.
}
\label{CH}
\end{center}
\end{figure}

Figures \ref{CT}(a)--\ref{CT}(c) show the temperature dependence of $C$ under various magnetic fields 
parallel to the $[001]$ ($\phi_B=0^\circ$), $[111]$ ($\phi_B=54.7^\circ$), and $[110]$ ($\phi_B=90^\circ$) axes, respectively.
In zero field, a steep increase in $C(T)$ is observed on cooling at $T_{\rm FQ} \sim 2.07$~K (midpoint).
By contrast, a specific-heat jump cannot be clearly observed at a superconducting transition temperature of 0.2~K~\cite{KittakaSM},
probably due to the residual magnetic field in the superconducting magnet (several mT) comparable to the upper critical field ($\sim 6$~mT)
and/or due to small electronic contribution to $C(T)$ in PrTi$_2$Al$_{20}$.
With increasing a magnetic field up to 5~T, the increase in $C(T)$ at $T_{\rm FQ}$ becomes gradual in $B \parallel [001]$ and $[110]$, whereas it remains nearly unchanged in $B \parallel [111]$.
These features are consistent with the previous reports~\cite{Sakai2011JPSJ,Sakai2012JPCS}.
In high magnetic fields, a prominent upturn is seen in $C(T)$ at low temperatures below roughly 0.4 K, which can be attributed to the nuclear contribution~\cite{KittakaSM}.

The broadening of the FQ transition can be more clearly seen in the field dependence of $C$.
Figures \ref{CH}(a) and \ref{CH}(b) show $C(B)$ measured at temperatures slightly above and below $T_{\rm FQ}$, respectively;
the increase (decrease) in $C(B)$ at 2.3 (1.8) K with increasing $B$ corresponds to the broadening of the FQ transition, as observed in the $C(T)$ data. 
Again, no clear change is detected in the $C(B)$ data in $B \parallel [111]$.
By contrast, the transition starts to broaden around 1~T for both $B \parallel [001]$ and $[110]$.
This indicates that the field-induced phase appears above roughly 1~T in these field orientations.
The broadening is more prominent in $B \parallel [001]$ than in $B \parallel [110]$.

A contrasting feature has been observed in $C(B)$ at low temperatures well below $T_{\rm FQ}$.
As shown in Fig.~\ref{CH}(c), 
$C(B)$ at 0.9~K is nearly insensitive to the magnetic field at least up to 5~T 
in $B \parallel [110]$ as well as $B \parallel [111]$,
whereas $C(B)$ decreases prominently in $B \parallel [001]$.
These features can also be confirmed in the $C(T)$ data (Fig.~\ref{CT}).
This field-insensitive behavior at low temperatures implies that 
the energy gap between the ground and first-excited states in the FQ ordered state does not change significantly with $B$.
This feature can be understood qualitatively in the framework of the conventional CEF model~\cite{KittakaSM}.

\begin{figure}[t]
\begin{center}
\includegraphics[width=2.75in]{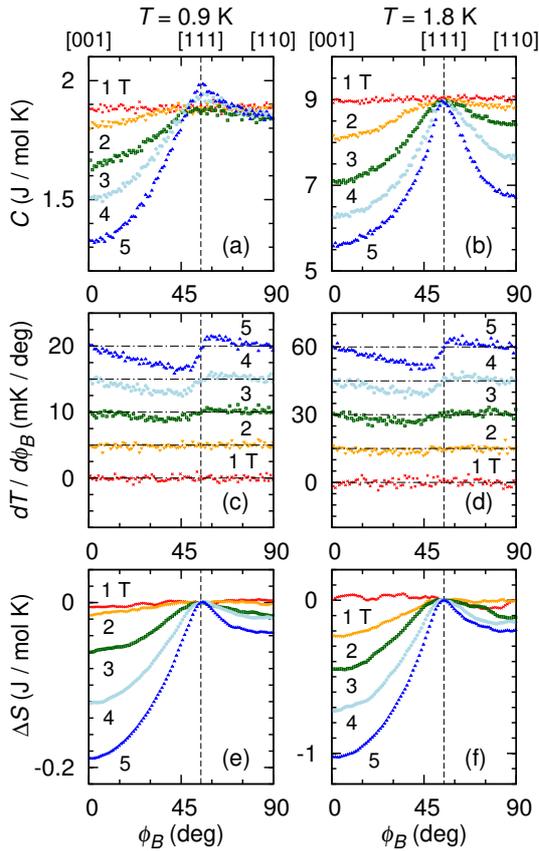}
\caption{
(Color online) Field-angle $\phi_B$ dependences of (a), (b) the specific heat, (c), (d) the rotational magnetocaloric effect, and 
(e), (f) relative change in the entropy measured from $S(\phi_B=54.7^\circ)$.
The data in (a), (c), and (e) [(b), (d), and (f)] were taken at 0.9 (1.8) K. Numbers labeling the curves are the applied magnetic fields.
Each set of the data in (c) and (d) are shifted vertically by 5 and 15 mK/deg, respectively, for clarity (dot-dashed lines represent $dT/d\phi_B=0$).
Dashed lines represent the $[111]$ direction. 
}
\label{Cf}
\end{center}
\end{figure}

To provide further insight into the field-orientation effect on the FQ order,
the $\phi_B$ dependence of $C$ has been measured at 0.9 and 1.8 K,
as shown in Figs. \ref{Cf}(a) and \ref{Cf}(b), respectively.
The high-field $C(\phi_B)$ exhibits a clear cusp at $\phi_B \sim 54.7^\circ$, i.e., in the $[111]$ direction.
At 1.8~K, slightly below $T_{\rm FQ}$, $C(\phi_B)$ smoothly decreases with rotating $B$ from $[111]$ to $[001]$ or $[110]$.
This result indicates that a small tilt of the magnetic-field orientation away from the $[111]$ axis transforms the FQ transition into a crossover
as the field strength becomes larger than 1~T.
No clear signature of a phase transition by the rotation of $B$ was detected from $C(\phi_B)$ measurements.

Figures \ref{Cf}(c) and \ref{Cf}(d) present the $\phi_B$ dependence of the rotational magnetocaloric effect at 0.9 and 1.8 K, respectively.
The field-rotational speed is set to $d\phi_B/dt=5$ and 2.5~s/deg at 0.9 and 1.8~K, respectively.
The data are obtained by averaging $dT/d\phi_B$ taken under clockwise and anticlockwise field rotations 
in order to cancel out the heat-transfer effect~\cite{Kittaka2018JPSJ}.
A clear rotational magnetocaloric effect is observed in high fields for $B \gtrsim 2$~T. 
It becomes zero when $B$ orients to the $[001]$, $[111]$, or $[110]$ direction.
This result suggests an occurrence of the temperature-independent zero-torque state ($\tau_\phi=0$) in these field orientations
because $dS/d\phi_B$ corresponds to $d\tau_\phi/dT$ according to the Maxwell relation.
Here, $\tau_\phi$ represents the magnetic torque within the field-rotation plane.

The $\phi_B$ dependence of the entropy can be evaluated 
as $S(\phi_B)=S(\phi_B=0^\circ)-\int^{\phi_B}_0C/T(\partial T/ \partial \phi_B)_{S,B}d\phi_B$
by using the results of specific-heat and rotational magnetocaloric effect measurements.
The relative entropy change defined as $\Delta S=S(\phi_B)-S(\phi_B=54.7^\circ)$ are plotted for $T=0.9$ and 1.8 K in Figs.~\ref{Cf}(e) and \ref{Cf}(f), respectively.
This definition is useful because the $C(T)$ data demonstrate that $S(\phi_B=54.7^\circ)$ is nearly unchanged with $B$ in the present range.
Whereas the entropy is strongly suppressed with increasing $B$ for $B \parallel [001]$, it is relatively robust against $B$ for $B \parallel [110]$.

\begin{figure}
\begin{center}
\includegraphics[width=2.75in]{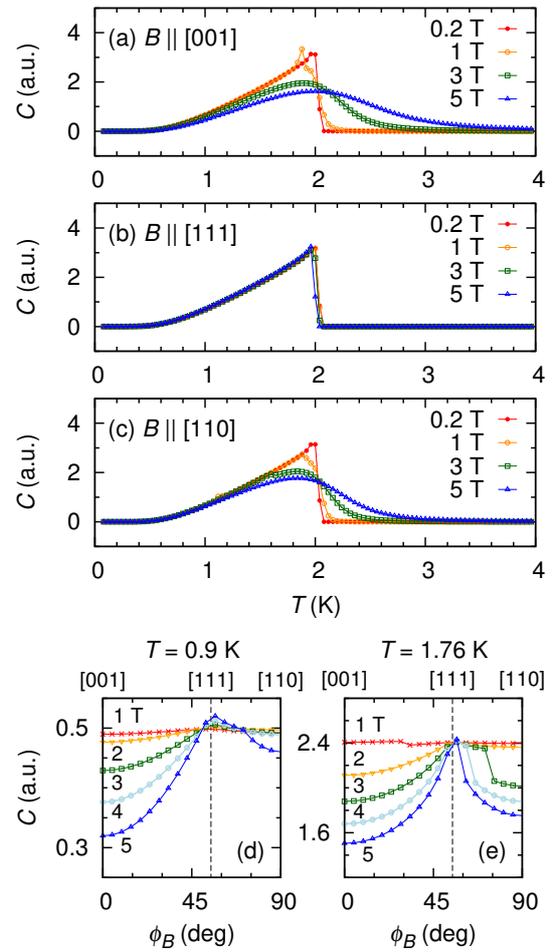}
\caption{
(Color online) 
Calculated results of the specific heat as a function of temperature for (a) $B \parallel [001]$, (b) $[111]$, and (c) $[110]$
and those as a function of the field angle $\phi_B$ at (d) 0.9 and (e) 1.76~K.}
\label{calc}
\end{center}
\end{figure}

In order to understand the observed field-angle-dependent phenomena,
$C(T,B,\phi_B)$ has been calculated by using the Hamiltonian within the $J=4$ multiplet of Pr$^{3+}$, 
\begin{equation}
\mathscr{H}=\mathscr{H}_{\rm CEF}+\mathscr{H}_{\rm Z}+\mathscr{H}_{\rm D}+\mathscr{H}_{\rm Q}+\mathscr{H}_{\rm O},
\end{equation}
which is the same one proposed in Ref.~\ref{Taniguchi2019}.
Here, the conventional CEF potential with the $T_d$ point symmetry $\mathscr{H}_{\rm CEF}$,
the Zeeman interaction $\mathscr{H}_{\rm Z}$, 
an exchange interaction between magnetic dipoles $\mathscr{H}_{\rm D}$,
a quadrupole-quadrupole interaction $\mathscr{H}_{\rm Q}$, and
a small $\Vec{T}^\alpha$ octupole-octupole interaction $\mathscr{H}_{\rm O}$ are expressed as
\begin{align}
\mathscr{H}_{\rm CEF}=& \epsilon_2|Q|^2-\epsilon_3\{Q_z^3-3\overline{Q_zQ_x^2}\}, \\
\mathscr{H}_{\rm Z}=&-g_J\mu_{\rm B}\Vec{J}\cdot\Vec{B},\\
\mathscr{H}_{\rm D}=&-\lambda_d\langle\Vec{J}\rangle\cdot\Vec{J},\\
\mathscr{H}_{\rm Q}=&-\lambda(\langle Q_z\rangle Q_z+\langle Q_x\rangle Q_x) \notag \\
&+\frac{\lambda^\prime f(|\Vec{B}|^2)}{2}[h_z(\langle Q_z\rangle Q_z - \langle Q_x\rangle Q_x) \notag \\
&-h_x(\langle Q_x\rangle Q_z+\langle Q_z\rangle Q_x)],\\
\mathscr{H}_{\rm O}=&-\lambda_T\langle \Vec{T}^\alpha\rangle \cdot \Vec{T}^\alpha,
\end{align}
respectively,
where $h_z=2B_z^2-B_x^2-B_y^2$ and $h_x=\sqrt{3}(B_x^2-B_y^2)$.
For convenience, two-component order parameters, i.e., $\Vec{Q}=(\langle Q_z\rangle, \langle Q_x\rangle)=Q(\cos\theta,\sin\theta)$, are adopted, 
where $Q_x=\sqrt{3}(J_x^2-J_y^2)/8$ ($\propto O_{22}$) and $Q_z=(3J_z^2-\Vec{J}^2)/8$ ($\propto O_{20}$).
The factor $f(|\Vec{B}|^2)=1/(1+c_2|\Vec{B}|^2+c_4|\Vec{B}|^4)$ is phenomenologically assumed so that
the effect of $\mathscr{H}_{\rm Q}$ ($\mathscr{H}_{\rm Z}$) becomes predominant at low (high) fields.
The parameters are set to the same as those in Ref.~\ref{Taniguchi2019}:
i.e., $\lambda=2$~K, $\lambda_d=-0.8$ K, $\lambda_T=0.0004$~K, $\lambda^\prime=0.08$~K/T$^2$, $c_2=0.8$~T$^{-2}$, and $c_4=0.8$~T$^{-4}$.

The calculated results of $C(T)$ for $B \parallel [001]$, $[111]$, and $[110]$ 
are plotted in Figs. \ref{calc}(a)--\ref{calc}(c), respectively.
Note that electronic and phonon contributions to the specific heat are not included in the present calculations.
The anisotropic broadening of the FQ transition as well as the low-temperature $C(T)$ insensitive to $B$ along the $[110]$ direction are reproduced reasonably.
Figures \ref{calc}(d) and \ref{calc}(e) present $C(\phi_B)$ calculated at 0.9 and 1.76~K ($<T_{\rm FQ}$), respectively, in a rotating field within the $(1\bar{1}0)$ plane.
Both the cusp structure in $C(\phi_B)$ at $\phi_B \sim 54.7^\circ$ and the gradual decrease in $C(\phi_B)$ with tilting $B$ from the $[111]$ to $[001]$ axes
are consistent with the experimental observations.

\begin{figure}[t]
\begin{center}
\includegraphics[width=2.75in]{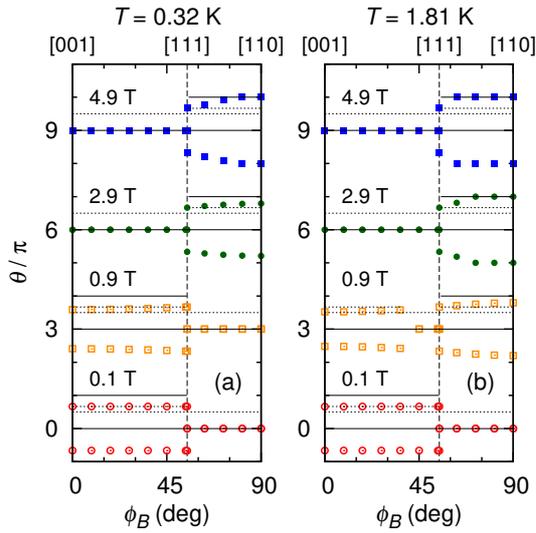}
\caption{
(Color online) 
Field-angle $\phi_B$ dependence of the order-parameter angle $\theta$ for (a) $T=0.32$ and (b) 1.81 K calculated at various magnetic fields.
Each set of the data is shifted vertically by 3 for clarity.
Long and short solid (dotted) lines represent $\theta=0$ and $\pi$ ($\theta=\pi/2$ and $2\pi/3$), respectively.
}
\label{theta}
\end{center}
\end{figure}

On the basis of the present model calculations, 
the occurrence of a phase transition has been predicted not only in the field scan but also in the temperature and field-angle scans.
For instance, a clear phase transition can be seen in $C(T)$ slightly below $T_{\rm FQ}$ at 1~T in $B \parallel [001]$ [Fig.~\ref{calc}(a)] and
in $C(\phi_B)$ around $\phi_B\sim75^\circ$ at 3~T and 1.76~K [Fig.~\ref{calc}(e)].
Although such anomalies have not been clearly detected from our specific-heat measurements,
the latter might be detected via $\Delta S(\phi_B)$, showing a small step-like change around $\phi_B \sim 75^\circ$ at 3~T and 1.8~K [Fig.~\ref{Cf}(f)]. 
This is probably because we directly measured the physical quantity corresponding to the field-angle derivative of the entropy, i.e., $(\partial T/\partial \phi_B)_{S,B}=-T/C(\partial S/\partial \phi_B)_{T,B}$, for determining $\Delta S(\phi_B)$.

As already reported in Ref.~\ref{Taniguchi2019}, 
the FQ order parameters with $\theta=0$ and $\pm 2\pi/3$ are stabilized in zero field or $B \parallel [111]$;
when $B \parallel [001]$ ($[110]$), $\pi/2\le|\theta|\le 2\pi/3$ ($\theta=0$) is stabilized in the low-field FQ phase and 
changes to $\theta=0$ ($2\pi/3\le|\theta|\le \pi$) in the field-induced phase.
Figures \ref{theta}(a) and \ref{theta}(b) show the order-parameter angle $\theta$ as a function of $\phi_B$ at 0.32 and 1.81~K, respectively, 
obtained from the present calculation.
The phase transition around $\phi_B\sim75^\circ$ at 3~T and 1.8~K 
can be attributed to a change in the order parameter from $2\pi/3<|\theta|< \pi$ to $|\theta|=\pi$;
in the latter phase, spontaneous symmetry breaking does not occur.

The main effect of the newly introduced $\lambda^\prime$ term is explained in the Supplemental Material \cite{KittakaSM}.
It is noted that a finite $\lambda^\prime$ is necessary to reproduce the low-field experimental data of $C(\phi_B)$ near $B \parallel [001]$.
Thus, the recently-proposed model, phenomenologically assuming an unconventional anisotropic quadrupole-quadrupole interaction, 
is in good agreements with the experimental observations under a rotating $B$ for PrTi$_2$Al$_{20}$.

In conclusion,
we have investigated the specific heat and entropy of PrTi$_2$Al$_{20}$
in a rotating magnetic field within the $(1\bar{1}0)$ plane.
It was found that the FQ transition becomes a crossover in fields larger than 1~T in any direction except for $B \parallel [111]$.
Our experimental results support the scenario
that the competition between the anisotropic quadrupole-quadrupole interaction and the Zeeman effect causes 
a discontinuous change in FQ order parameters under a magnetic field.
The field-angle-resolved thermal measurements can be a useful tool to test order parameters for multiple phases.

\acknowledgments
This work was supported by a Grants-in-Aid for Scientific Research on Innovative Areas ``J-Physics'' (15H05882, 15H05883, 18H04306, 18H04310)
from MEXT, and KAKENHI (18H01161, 18H01164, 18K03522, 17H02918, 17J08806, 19H00650) from JSPS.

\bibliographystyle{jpsj2}
\bibliography{C:/usr/local/share/texmf/bibref/ref_PrTi2Al20.bib}

\clearpage
\normalsize
\onecolumn
\renewcommand{\thefigure}{S\arabic{figure}}
\setcounter{figure}{0}

\begin{center}
{\large Supplemental Material for \\
\textbf{Field-Orientation Effect on Ferro-Quadrupole Order in PrTi$_2$Al$_{20}$}}\\
\vspace{0.1in}
Shunichiro Kittaka,$^{1}$ Takanori Taniguchi,$^1$ Kazumasa Hattori,$^2$ Shota Nakamura,$^{1}$ Toshiro Sakakibara,$^{1}$ 
Masashi Takigawa,$^1$ Masaki Tsujimoto,$^1$ Akito Sakai,$^1$ Yosuke Matsumoto,$^1$ Satoru Nakatsuji$^{1,3}$\\
\vspace{0.1in}
{\small 
\textit{$^1$Institute for Solid State Physics, University of Tokyo, Kashiwa, Chiba 277-8581, Japan}\\
\textit{$^2$Department of Physics, Tokyo Metropolitan University, Hachioji, Tokyo, 192-0937, Japan}\\
\textit{$^3$Department of Physics, University of Tokyo, Hongo, Bunkyo-ku, Tokyo 113-0033, Japan}\\
}
\end{center}

\normalsize
\vspace{0.2in}
\textbf{\large I.\ \  Low-temperature specific heat of PrTi$_2$Al$_{20}$}
\vspace{0.1in}

The low-temperature part of $C(T)/T$ for $B \parallel [111]$ is shown in Fig. S1.
Although the ferroquadrupole transition at $T_{\rm FQ} \sim 2$~K is nearly unchanged in this field orientation, 
the low-temperature upturn in $C/T$ becomes prominent with increasing $B$. 
This upturn can be fitted satisfactorily by using a function $f(T)=a+b/T^3$, as represented by dashed lines in Fig. S1.
The fitting parameter $b$ is plotted as a function of $B$ in the inset of Fig. S1(a).
It is nearly proportional to $B^2$, which is reminiscent of the nuclear Schottky contribution.

Let us calculate the nuclear specific heat $C_{\rm N}(T,B)=b_{\rm calc}(B)/T^2$ of PrTi$_2$Al$_{20}$.
Here, the effect of a quadrupole splitting of nuclei is ignored; 
indeed, the upturn in $C/T$ in zero field is small and does not follow $T^{-3}$ dependence [Fig. S1(b)].
The observed $C_{\rm N}$ is mostly caused by $^{27}$Al ($I=5/2$ with the natural abundance of 100\%) and 
$^{141}$Pr ($I=5/2$ with the natural abundance of 100\%) nuclei.
The former contribution can be evaluated as $b_{\rm calc}^{\rm Al}(B)=0.14B^2$~mJ K /mol
by using a nuclear spin Hamiltonian;
\begin{equation}
\mathscr{H}_{\rm Al}=-g_{\rm N}^{\rm Al}\mu_{\rm N}\Vec{I}\cdot\Vec{B},
\end{equation}
where $g_{\rm N}^{\rm Al}=1.46$ is the nuclear $g$ factor of an Al ion and $\mu_{N}$ is the nuclear magneton.

Because a strong hyperfine coupling arises between a Pr nucleus and $4f$-electrons on the same ion,
contribution from Pr nuclei can be calculated by using the Hamiltonian~\cite{Aoki2011JPSJS}
\begin{equation}
\mathscr{H}_{\rm Pr}=\Vec{I}\cdot(A_{\rm hf}\Vec{J}-g_{\rm N}^{\rm Pr}\mu_{\rm N}\Vec{B}),
\end{equation}
where $A_{\rm hf}$ is a coupling constant of hyperfine interaction of a Pr ion,
$\Vec{J}$ is the total angular momentum of $4f$ electrons, and $g_{\rm N}^{\rm Pr}=1.71$ is the nuclear $g$ factor of a Pr ion~\cite{Macfarlane1982PRLS}.
Then, the latter contribution $b_{\rm calc}^{\rm Pr}(B)$ can be expressed by
\begin{equation}
b_{\rm calc}^{\rm Pr}(B)=R[A_{\rm hf}m_{\rm Pr}(B)/g_{J}+g_{\rm N}^{\rm Pr}\mu_{\rm N}B]^2I(I+1)/(3k_{\rm B}^2),
\end{equation}
where $R$ is the gas constant, $m_{\rm Pr}=g_{J}(\overline{|J_z|^2})^{1/2}$ is the site-averaged magnitude of the Pr magnetic moment, $g_J=4/5$ is the Land$\acute{\rm e}$ $g$ factor, and $k_{\rm B}$ is the Boltzmann constant.
In this calculation, we adopt $A_{\rm hf}/k_{\rm B}=0.052$~K~\cite{Kondo1961JPSJS} and $m_{\rm Pr}(B)=0.068B$ $\mu_{\rm B}$/Pr for $B \le 5$~T~\cite{Taniguchi2019JPSJS}.
Then, we obtain $b_{\rm calc}^{\rm Pr}(B)=0.62B^2$~mJ K /mol. 
The sum of these nuclear contributions is $b_{\rm calc}(B) = b_{\rm calc}^{\rm Al}(B) + b_{\rm calc}^{\rm Pr}(B) = 0.76B^2$ mJ K / mol [a solid line in the inset of Fig.~S1(a)].
Good agreement between experimental and calculated results demonstrates that
the observed upturn in $C/T$ can be attributed to the nuclear contribution.

Figure S1(b) shows the low-temperature $C(T)/T$ in zero field.
In a previous report \cite{Sakai2012JPSJS}, the occurrence of a superconducting transition was reported at $T_{\rm c}=0.2$~K.
A slight deviation from the $T^{-3}$ behavior is seen below 0.2 K in Fig. S1(b),
which might be attributed to the superconducting transition.
However, it is rather broad and other possible origins, e.g., nuclear contribution, cannot be fully ruled out.
Because the upper critical field is roughly 6~mT, 
a residual magnetic field in a superconducting magnet (typically several mT) would suppress the superconducting transition.

\vspace{0.3in}
\textbf{\large II.\ \  Effect of the $\lambda^\prime$ term in the calculation}
\vspace{0.1in}

Figure S2 shows the calculated results of $C(\phi_B)$ with (a), (b) $\lambda^\prime=0$ and (c), (d) $\lambda^\prime=0.08$ K/T$^2$.
The experimental results of $C(\phi_B)$ in high fields can be reproduced qualitatively from the calculation with $\lambda^\prime=0$ as well.
The $\lambda^\prime$ term mainly affects the low-field data in $C(\phi_B)$ around $B \parallel [001]$.
For instance, at 0.9 K, the difference between $C(\phi_B)$ at 1 and 2~T becomes small with changing $\lambda^\prime$ from 0 to 0.08~K/T$^2$.
Moreover, at 1.76~K, $C(\phi_H)$ in $B \parallel [001]$ is comparable to that in $B \parallel [111]$ when $\lambda^\prime=0.08$~K/T$^2$,
whereas $C(\phi_H)$ gradually decreases with rotating $B$ from the $[111]$ to $[001]$ axes for $\lambda^\prime=0$.
Thus, the calculated results for $\lambda^\prime=0.08$ K/T$^2$ provide a better match with the experimental observations in $C(\phi_B)$ and $S(\phi_B)$ at low fields.

\clearpage
\renewcommand{\thefigure}{S\arabic{figure}}
\setcounter{figure}{0}

\begin{figure}
\begin{center}
\includegraphics[width=3.3in]{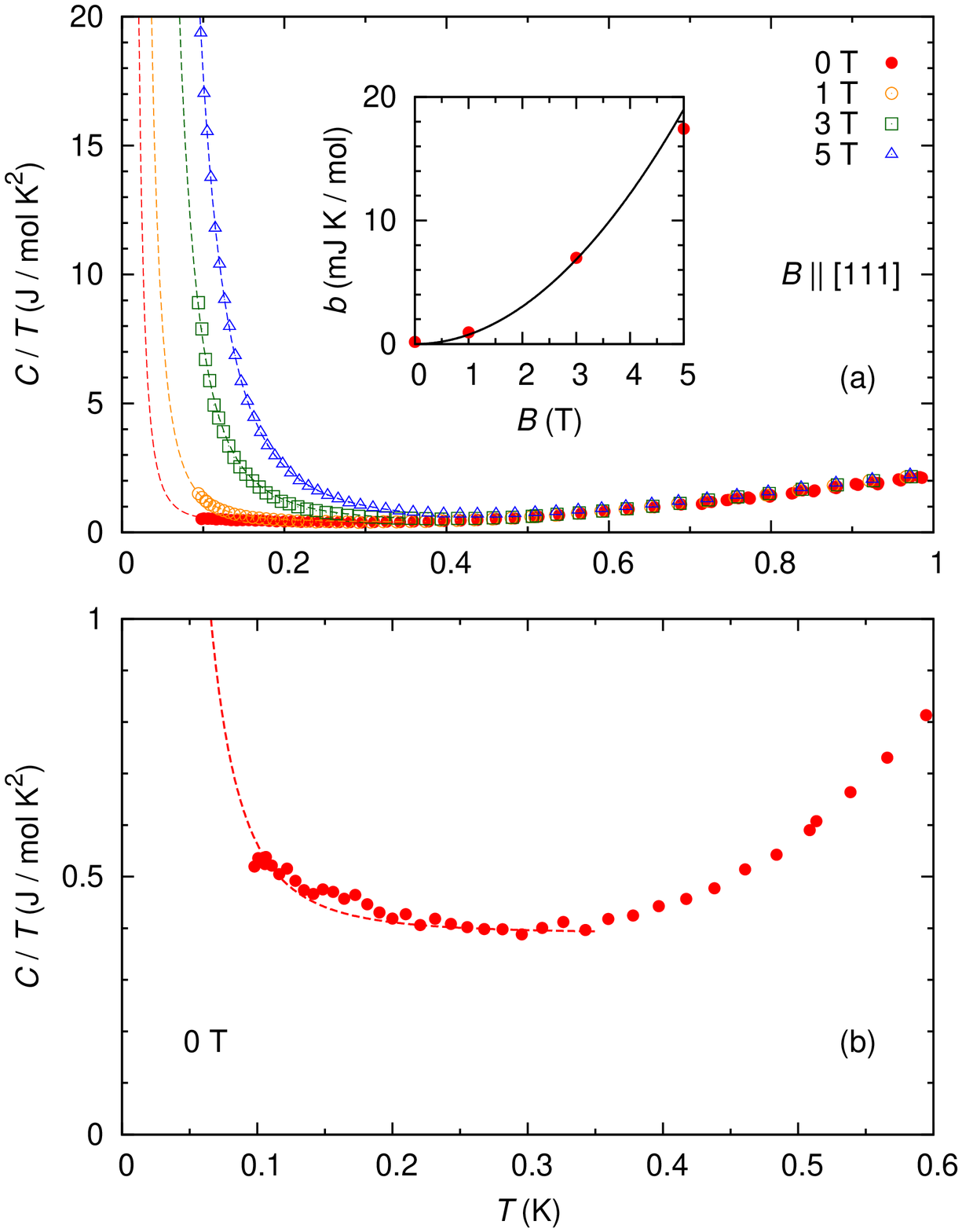} 
\end{center}
\caption{
(a) Temperature dependence of $C/T$ at several magnetic fields in $B \parallel [111]$. 
Dashed lines represent fits to the data below 0.35~K using the function $f(T)=a+b/T^3$.
The inset shows $b(B)$ obtained by the fits (circles). A solid line represents $b_{\rm calc}(B)$ obtained from the calculation using the nuclear spin Hamiltonian (see text).
(b) The low-temperature $C(T)/T$ in zero field.
}
\label{CH001}
\end{figure}

\begin{figure}
\vspace{0.2in}
\begin{center}
\includegraphics[width=3.3in]{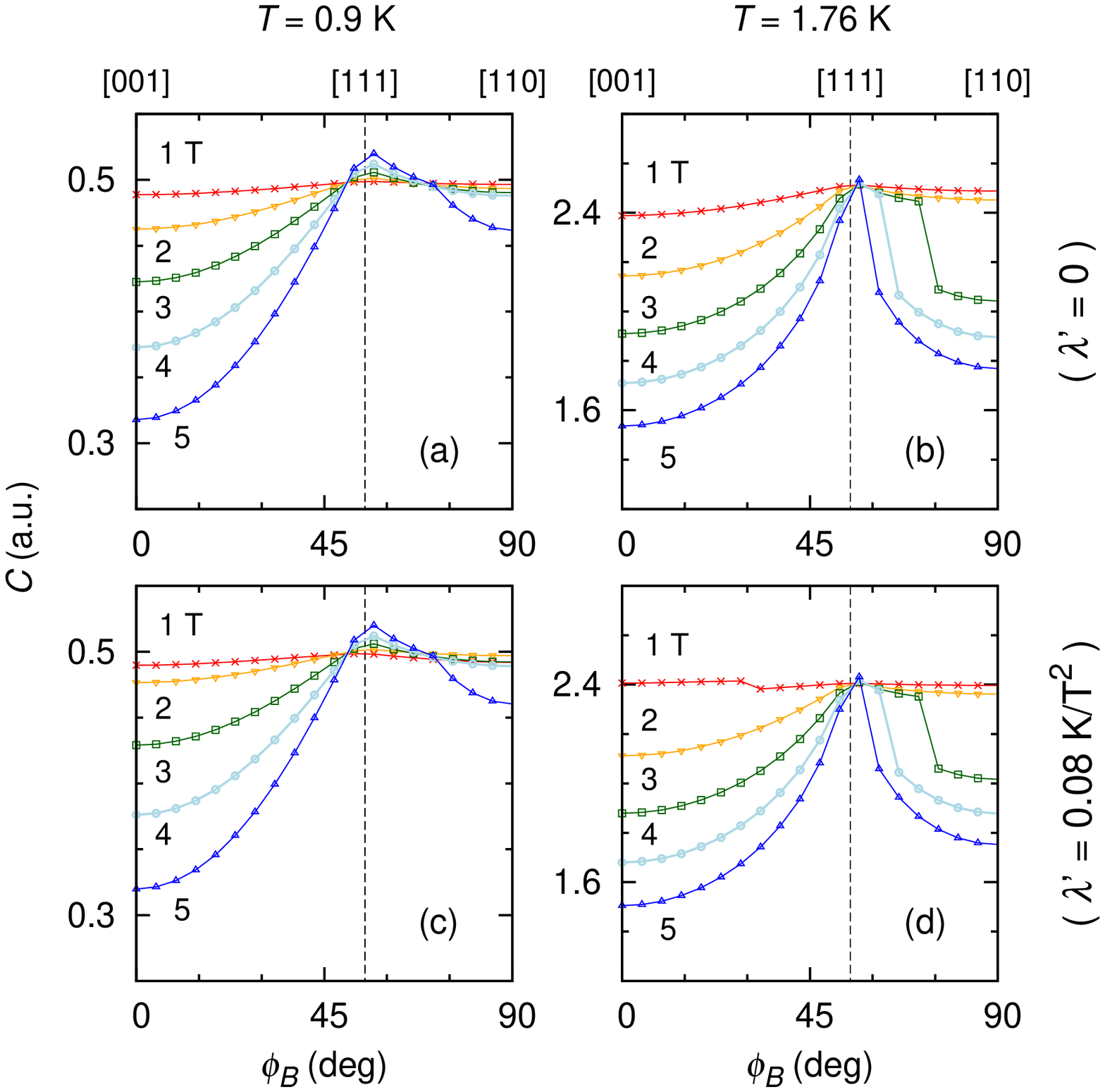} 
\end{center}
\caption{
Calculated results of the specific heat as a function of the field angle $\phi_B$ at (a) [(c)] 0.9 and (b) [(d)] 1.76~K with $\lambda^\prime=0$ (0.08) K/T$^2$.
}
\label{CH001}
\end{figure}

\end{document}